\documentclass[graphics,floatfix, footinbib,tightenlines,nobibnotes, aps, prl, twocolumn]{revtex4-1}
\usepackage{amsmath,amssymb,wasysym}
\usepackage{graphicx}% Include figure files
\usepackage{dcolumn}% Align table columns on decimal point
\usepackage{bm}% bold math
\usepackage{braket}
\usepackage{verbatim}
\usepackage{float}
\usepackage{bbold}
\usepackage{color}
\usepackage{xcolor}
\usepackage{relsize}
\usepackage{amsthm}
\usepackage{enumerate}
\usepackage{soul,xcolor}
\usepackage[T1]{fontenc}
\usepackage{adjustbox}
\usepackage{hyperref}
\hypersetup{
colorlinks = true,
linkcolor = [rgb]{0.70,0.13,0.13},
citecolor = [rgb]{0.13,0.55,0.13},
urlcolor = [rgb]{0.25, 0.41, 0.88}}

\newcommand{\be}{\begin{equation}}
\newcommand{\ba}{\begin{align}}
\newcommand{\ee}{\end{equation}}
\newcommand{\bea}{\begin{eqnarray}}
\newcommand{\eea}{\end{eqnarray}}
\newcommand{\beq}{\begin{equation}}
\newcommand{\eeq}{\end{equation}}
\newcommand{\beqn}{\begin{eqnarray}}
\newcommand{\eeqn}{\end{eqnarray}}

\definecolor{Red}{rgb}{0.70,0.13,0.13}

\usepackage{bm}

\usepackage{array}
\newcolumntype{L}[1]{>{\raggedright\arraybackslash}p{#1}}
\newcolumntype{C}[1]{>{\centering\arraybackslash}p{#1}}
\newcolumntype{R}[1]{>{\raggedleft\arraybackslash}p{#1}}
\usepackage{multirow}

\allowdisplaybreaks

\begin{document}

%\title{Quantum simulation of high T$_c$ nickelate superconductor in tetralayer optical lattice}

\title{Strong pairing and symmetric pseudogap metal in double Kondo lattice model: from nickelate superconductor to tetralayer optical lattice}

\author{Hui Yang}
\thanks{These two authors contributed equally}
\author{Hanbit Oh}
\thanks{These two authors contributed equally}
\author{Ya-Hui Zhang}
\affiliation{William H. Miller III Department of Physics and Astronomy, Johns Hopkins University, Baltimore, Maryland, 21218, USA}

\date{\today}

\begin{abstract}
In this work, we propose and study a double Kondo lattice model which hosts robust superconductivity. The system consists of two identical Kondo lattice model, each with Kondo coupling $J_K$ within each layer, while the localized spin moments are coupled together via an inter-layer on-site antiferromagnetic spin coupling $J_\perp$.  We consider the strong $J_\perp$ limit, wherein the local moments tend to form rung singlets and are thus gapped. However, the Kondo coupling $J_K$ transmits the inter-layer entanglement between the local moments to the itinerant electrons. Consequently, the itinerant electrons experience a strong inter-layer antiferromangetic spin coupling and form strong inter-layer pairing, which is confirmed through numerical simulation in one dimensional system. Experimentally, the $J_K \rightarrow -\infty$ limits of the model describes the recently found bilayer nickelate La$_3$Ni$_2$O$_7$, while the $J_K>0$ side can be realized in tetralayer optical lattice of cold atoms.  
Two extreme limits,  $J_K \rightarrow -\infty$ and $J_K \rightarrow +\infty$ limit are shown to be simplified to a bilayer type II t-J model and a bilayer one-orbital t-J model, respectively. 
Thus, our double Kondo lattice model offers a unified framework for nickelate superconductor and tetralayer optical lattice quantum simulator upon changing the sign of $J_K$. We highlight both the qualitative similarity and the quantitative difference in the two sides of $J_K$. Finally, we discuss the possibility of a symmetric Kondo breakdown transition in the model with a symmetric pseudogap metal corresponding to the usual heavy Fermi liquid.
 \end{abstract}

\maketitle

\textit{Introduction}: Kondo lattice model is one of the most important strongly correlated models in condensed matter physics and describes a variety of interesting phenomena in heavy fermion systems, including heavy Fermi liquid, superconductivity and quantum criticality\cite{coleman2001fermi,gegenwart2008quantum,si2010heavy,stewart2001non,coleman2005quantum,lohneysen2007fermi,senthil2005quantum,kirchner2020colloquium}. In the conventional spin-$\frac{1}{2}$ Kondo lattice model, itinerant electron couples to localized spin-half moments with the Kondo coupling $J_K$. Depending on the competition between $J_K$ and the spin-spin interaction within the  local moments, the ground state is in a heavy Fermi liquid or a Kondo breakdown phase. At intermediate $J_K$ there may be a superconductor dome and a quantum critical regime\cite{friedemann2009detaching,trovarelli2000ybrh,schroder2000onset} with strange metal behavior. It was proposed that the transition in the Kondo lattice model may be associated with a jump in Fermi surface volume resulting from Kondo breakdown, rather than from symmetry-breaking orders as in the Herts-Millis theory\cite{hertz1976quantum,millis1993effect}. Despite theoretical efforts using various different methods\cite{si2001locally,senthil2004weak,senthil2003fractionalized,paul2007kondo,zhang2020pseudogap,zhang2020deconfined}, a well-established theory of the Kondo breakdown transition is still elusive. One particular difficulty is that the local moments usually form  magnetic order in the Kondo breakdown phase. As a result, the transition needs to incorporate both the Kondo breakdown and the onset of the magnetic ordering simultaneously.  Actually the metal in the small $J_K$ side is smoothly connected to a conventional symmetry breaking Fermi liquid, thus it is not even clear that the transition is necessarily beyond the Landau-Ginzburg framework.

In this work, we introduce a new model dubbed as double Kondo lattice model. The model consists of two layers of the conventional spin-half Kondo lattice model with Kondo coupling $J_K$. Then we add an on-site inter-layer antiferromagnetic spin-spin coupling $J_\perp$ between the local moments in the two layers. The $J_\perp=0$ limit reduces to two decoupled conventional Kondo lattice model. Here we are interested in the regime of strong $J_\perp>0$. With a large $J_\perp$, the localized spin moments tend to form rung singlets and get gapped out. However, the Kondo coupling $J_K$ transmits the inter-layer entanglement between the local moments to the itinerant electrons.  As a result, there is an effective anti-ferromagnetic spin coupling $\tilde J_\perp$ between the itinerant electrons even though the localized moments get gapped. One consequence is that the ground state is an inter-layer paired superconductor for both $J_K>0$ and $J_K<0$ sides. The strong pairing persists in a wide range of $|J_K|$, in contrast to the conventional single layer Kondo lattice model where superconductor dome is usually restricted to the critical regime at intermediate $J_K$. In addition to the robust superconductor phase at low temperature, another attractive feature of the model is that there are two different symmetric Fermi liquids in the normal state above the superconducting critical temperature ($T_c$). These two Fermi liquids correspond to the Kondo breakdown phase and the heavy Fermi liquid. In this new model, the Kondo breakdown phase is just the usual Fermi liquid (FL) phase decoupled with a trivial rung-singlet from the local spin moments. On the other hand, the heavy Fermi liquid(HFL) has a different Fermi surface volume. If we only look at the itinerant electrons given that the localized spin moments are gapped, the heavy Fermi liquid can be viewed as a symmetry pseudogap metal with small Fermi surfaces, akin to certain candidates of the pseudogap phase in underdoped cuprate\cite{zhang2020pseudogap}. Therefore the normal states in small and large $|J_K|$ must be separated by a phase transition, but without any symmetry breaking, offering a clean model to study quantum criticality beyond Landau-Ginzburg theory.

We also propose two potential experimental systems to realize the double Kondo model with $J_K<0$ and $J_K>0$ respectively. First, we suggest that the negative $J_K$ side of the double Kondo lattice model describes the  recently observed  nickelate superconductor in the bilayer La$_{3}$Ni$_{2}$O$_{7}$ under high pressure with $T_{c} = 80$K \cite{sun2023signatures}. 
Following the experimental observation, extensive studies, both theoretical\cite{luo2023bilayer,zhang2023electronic,yang2023possible,sakakibara2023possible,gu2023effective,shen2023effective,wu2023charge,christiansson2023correlated,liu2023s,cao2023flat,qu2023bilayer,lu2023superconductivity,jiang2023pressure,tian2023correlation,zhang2023strong,qin2023high,huang2023impurity,zhang2023trends,jiang2023high,yang2023minimal,qu2023bilayer,2023arXiv230809044Q,2023arXiv230807386Z,2023arXiv230812750K,2023arXiv230811614J,chen2024non,PhysRevB.109.L201124,zhan2024cooperation} and experimental\cite{liu2023electronic,hou2023emergence,zhang2023high,yang2023orbital,zhang2023effects,Dong2024}, have been done to unravel the underlying physics of the novel pairing mechanism in the bilayer nickelate. 
Especially, the  Hund’s coupling between the $d_{x^2-y^2}$ orbital and the $d_{z^2}$ orbital has been shown to play a key role in achieving the remarkably high $T_c$ \cite{oh2023type,lu2023interlayer}. As is proposed by us previously\cite{oh2023type,yang2023strong,2024arXiv240500092O}, the $d_{z^2}$ orbital is Mott localized and only provides a spin-1/2 moment per site. The $d_{x^2-y^2}$ orbital provides itinerant electrons which couple to the localized moments through the on-site Hund's coupling $J_H$. Then the physics is exactly captured by the double Kondo lattice model with $J_K=-2J_H<0$. On the other hand, the $J_K>0$ side of the double Kondo model can be realized in tetralayer optical lattice, with each layer hosting a Fermi-Hubbard model\cite{RevModPhys.80.885}. By introducing a potential bias $\Delta<U$ between the inner and outer layers, we can simulate the double  Kondo lattice model, where the conduction electrons are residing on the outer two layers and the localized moments live on the inner two layers, as illustrated in Fig \ref{fig:1}. 
Notably, the Kondo coupling $J_K$ is now from the super-exchange and is always positive. $J_K$ can be conveniently tuned by either inter-layer hopping or potential difference, providing a wonderful controllable platform to study the rich physics in the model. In summary, the double Kondo lattice model offers a unified framework for the bilayer nickelate and the tetralayer optical lattice with different signs of $J_K$. We will see that the two sides share qualitatively similar physics, but the positive $J_K$ side achieves significantly stronger pairing.

\begin{figure}[tb]
    \centering
\includegraphics[width=1.00\linewidth]{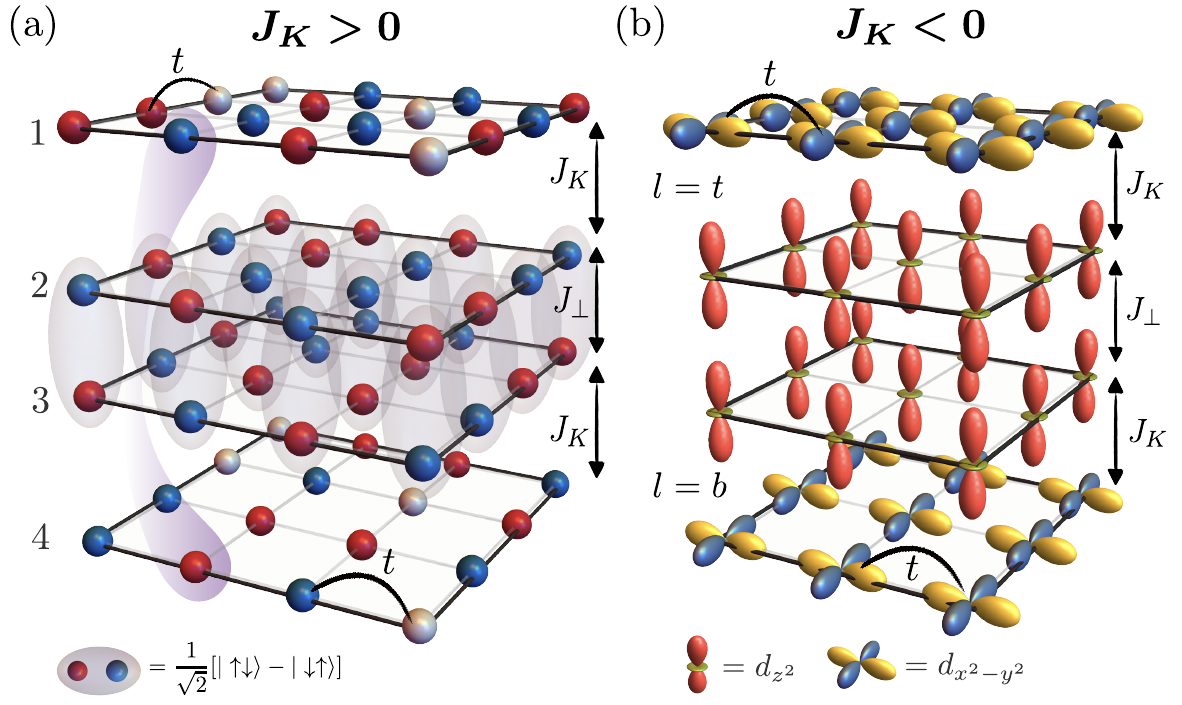}
\caption{\textbf{Illustration of the tetralayer  optical lattices to realize the double Kondo lattice model on square lattice and bilayer nickelate.}
(a) Tetralayer optical lattice simulation of the double Kondo lattice model using the ultracold atoms. 
The potential offset $\Delta$ between $1\leftrightarrow 2$ ($3\leftrightarrow 4$) enforces the filling, $n_1=n_4=1-x$ and $n_2=n_3=1$. In this case $J_K>0$ is from the super-exchange.
(b) In the bilayer nickelate, the $d_{x^2-y^2}$ offers itinerant electrons with an intra-layer hopping $t$ but no inter-layer hopping. Meanwhile, the $d_{z^2}$ orbital is Mott localized and provides a spin 1/2 local moment per site. Here we separate the $d_{x^2-y^2}$ and $d_{z^2}$ orbitals to two layers just for illustration. In reality they are in the same site and coupled together by the on-site coupling $J_K=-2J_H<0$. 
The double Kondo lattice model in $J_{K}\rightarrow +\infty$ reduces to the bilayer $t-J-J_{\perp}$ model, while the $J_{K}\rightarrow -\infty$ limit reduces to the type-II t-J model. We emphasize that the $J_K\rightarrow \pm \infty$ limits are different with distinct Hilbert spaces. The previous proposals of using the bilayer $t-J-J_\perp$ model (suitable for $J_K\rightarrow +\infty$) to describe the bilayer nickelate (with $J_K<0$) is not appropriate.
    }
    \label{fig:1}
\end{figure}

\textit{Double Kondo lattice model}:
We start with the Hubbard model on a four-layer optical lattice (See Fig. \ref{fig:1} (a)). We add a potential difference between the inner two layers compared to the outer two layers so the density is $n_1=n_4=1-x, n_2=n_3=1$ for the four layers respectively. Due to the large Hubbard $U$, the electrons in the inner two layers are Mott localized and only providing localized spin moments. We label the inner two layers as S layers and the outer two layers as C layers.  We can then split the four layers to two groups labeled as top and bottom. Each group consists of a C layer and a S layer coupled together by a Kondo coupling $J_K$.  For the S layer, 
we label the localized spin operator at layer 2 (3) as $\vec S_{i;t}$ ($\vec S_{i;b}$). Here, $t(b)$ is an acronym of top (bottom) layer, and  $i$ is a site index. 
Similarly, for the C layer, 
we use $c_{i;t \sigma}$ ($c_{i;b \sigma}$) to indicate electron operators at layer  1 (4) carrying a spin number $\sigma=\uparrow,\downarrow$, respectively. 
Then, the spin operator of the electron at the C layers defined as $\vec s_{c;i;a}=\frac{1}{2} \sum_{\sigma,\sigma'}c^\dagger_{i;a\sigma} \vec \sigma_{\sigma \sigma'}c_{i;a\sigma'}$.

Performing $t/U$ expansion, we can reach  the double Kondo lattice model, 
\begin{align}
 H=&
 -t\sum_{a=t,b} \sum_{\langle i,j \rangle}
Pc^\dagger_{i;a;\sigma} c_{j;a;\sigma}P+J_{c}
\sum_{a=t,b}
\sum_{\langle i,j \rangle }
\vec s_{c;i;a} \cdot \vec s_{c;j;a}
\nonumber
\\
&
+J_{K}
\sum_{i}\sum_{a=t,b} \vec s_{c;i;a} \cdot \vec S_{i;a}+J_{s}
\sum_{a=t,b}
\sum_{\langle i,j \rangle }
\vec S_{i;a} \cdot \vec S_{j;a}
\nonumber
\\
& +J_{\perp}\sum_{i}
\vec S_{i;t}\cdot \vec S_{i;b}
\label{eq:Ham_optical}
\end{align}
where $P$ is the projection operator to remove the double occupancy at each site in the C layers. The first two lines describe two decoupled Kondo lattice model \cite{zhang2022pair} for $a=t,b$ respectively. Note that the itinerant electron is itself described by a t-J model. The $J_\perp$ term couples the localized spin moments of the two Kondo models together. In this work, our scope of interests is restricted in the large and positive $J_\perp$ regime. Both $J_K$ and $J_\perp$ terms originate from the super-exchange and can be conveniently controlled by the inter-layer hopping or the potential offset (see Supplemental Material (SM) Sec.\textcolor{Red}{I}).

In addition, the same double Kondo lattice model can also be realized in bilayer nickelate, as illustrated in Fig.\ref{fig:1}(b). 
Actually the model was proposed by us to describe the recent observed $80$ K high temperature superconductivity in the pressure La$_3$Ni$_2$O$_7$ system \cite{oh2023type,yang2023strong}. In bilayer nickelate, the $d_{x^2-y^2}$ and $d_{z^2}$ orbitals of the Ni atom play the role as the C layer and S layer in the double kondo model. The Kondo coupling $J_K$ is now from the on-site Hund's coupling $J_H>0$ and we have $J_K=-2J_H<0$.  But $J_\perp$ is still positive and large due to large inter-layer hopping of the $d_{z^2}$ orbital.

\textbf{Type II t-J model in the $J_K \rightarrow - \infty$ limit}: In the negative and large $J_K$ limit, we can first solve the $J_K$ term at each site $i$ for the $a=t,b$ group, ignoring the intersite hopping and the inter group coupling.
The original local Hilbert space at every site $i$ and layer $a$ is $3\times 2=6$ dimensional, which is from a tensor product of the three states in the C layer and the two states in the S layer.
However, in $J_K\rightarrow -\infty$ limit, only the $5=3+2$ states are active in low energy because the spin-singlet is forbidden by the large energy penalty $-J_K$.
The $5$ states can be grouped to two categories: two $S=\frac{1}{2}$ singlon states with $n_{i;a}=0$ and three $S=1$ doublon states with $n_{i;a}=1$.  The two singlon states are written as $\ket{0}_C \otimes \ket{\sigma}_S$, where we define $\ket{0}_C$ as the empty state of the C layer and $\ket{\sigma}_S$ as the localized spin state of the S layer with $\sigma=\uparrow,\downarrow$. 
The three doublon states are defined as the spin-triplet state, $\ket{t}_1=\ket{\uparrow}_C\otimes\ket{\uparrow}_S$, $\ket{t}_0=\frac{1}{\sqrt{2}}[\ket{\uparrow}_C\otimes\ket{\downarrow}_S+\ket{\downarrow}_C\otimes\ket{\uparrow}_S]$, $\ket{t}_{-1}=\ket{\downarrow}_C\otimes\ket{\downarrow}_S$. 
By projecting the original Hamiltonian in the restricted Hilbert space with these five states, one can reach a bilayer type II t-J model as discussed in Refs.~\cite{zhang2020type,oh2023type,yang2023strong},
\begin{align}
     H=
     -&t\sum_{a=t,b}
\sum_{\langle i,j \rangle}
\sum_{\sigma}
\left[Pc^{\dagger}_{i;a;\sigma}
c_{j;a;\sigma}P
+h.c.\right], 
     \notag\\
     +&  \sum_{a=t,b} \sum_{\langle i,j \rangle}
   \left[
   J^{ss}_\parallel
\vec S_{i;a}\cdot \vec S_{j;a}
      +
      J^{dd}_\parallel
\vec T_{i;a}\cdot \vec T_{j;a} \notag 
      \right]
      \notag\\
+& \sum_{a=t,b} \sum_{\langle i,j \rangle}J^{sd}_\parallel
      \left[\vec S_{i;a}\cdot \vec T_{j;a} +\vec T_{i;a}\cdot \vec S_{j;a}\right]
      \notag\\
+& \sum_{i} \left[J^{ss}_\perp\vec S_{i;t}\cdot \vec S_{i;b}
+J ^{dd}_\perp  \vec T_{i;t}\cdot \vec T_{i;b} \right],\notag\\
+& \sum_{i} J^{sd}_\perp
\left[\vec S_{i;t}\cdot \vec T_{i;b}+\vec T_{i;t}\cdot \vec S_{i;b}\right],
\label{eq:type_II_t_J}
\end{align}
with $J^{ss}_{\perp}=2J^{sd}_{\perp}=4J^{dd}_{\perp}=J_{\perp}$, and 
$J^{ss}_{\parallel}=J_s$, $J^{sd}_{\parallel}=J_s/2$, $J^{dd}_{\parallel}=J_{c}/4+J_{s}/4$. In the above $c_{i;a\sigma}$ is the projected electron operator. $\vec{S}_{i,a}=\frac{1}{2}\sum_{\sigma\sigma'=\uparrow,\downarrow} \vec{\sigma}_{\sigma\sigma'}\ket{\sigma}_{i,a}\bra{\sigma'}_{i,a}$ is the spin-1/2 operator, and 
$ \vec T_{i,a}=\sum_{\alpha,\beta=-1,0,1} \vec T_{\alpha \beta}\ket{t}_{i,a;\alpha} \bra{t}_{i,a;\beta}$ is the spin-1 operator. The details on the bilayer type-II t-J model is provided in SM Sec.\textcolor{Red}{II}).

\textbf{One orbital t-J model in the $J_K \rightarrow +\infty$ limit}: 
In the positive and large $J_K$ limit, the situation is different, as the spin-singlet doublon state now has lower energy.
More specifically, we only have one $S=0$ doublon state (defined as $n_{i;a}=1$): $\ket{s}=\frac{1}{\sqrt{2}}(\ket{\uparrow}_C\otimes\ket{\downarrow}_S-\ket{\downarrow}_C\otimes\ket{\uparrow}_S)$. 
There are still two $S=\frac{1}{2}$ singlon states: $\ket{0}_C \otimes \ket{\sigma}_S$. 
Therefore,  the Hilbert space is reduced to b $3=1+2$-dimensional for each group $a=t,b$.
Then, the singlet state $\ket{s}$ can be treated as the new empty state in the one-orbital t-J model. Adopting the same notation of the conventional t-J model, we define new creation operator $\tilde{c}^\dag_{i;a\sigma}=\ket{\sigma}_{i;a}\bra{s}_{i;a}$, and density operator $\tilde n_{i;a}=\sum_{\sigma}\ket{\sigma}_{i;a}\bra{\sigma}_{i;a}$. In the end, we can derive a bilayer one-orbital t-J model,
\begin{align}
    H=&
 -\tilde{t}\sum_{a=t,b} \sum_{\langle i,j \rangle}
\sum_{\sigma}
P\tilde{c}^\dagger_{i;a;\sigma} \tilde{c}_{j;a;\sigma}P+
J_{\perp}\sum_{i}
\vec S_{i;t}\cdot \vec S_{i;b}
\nonumber
\\
& +J_{\parallel}
\sum_{a=t,b}
\sum_{\langle i,j \rangle }
\vec S_{i;a} \cdot \vec S_{j;a},
\label{eq:bilayer_one_orbital}
\end{align}
with $\tilde{t}=-\frac{1}{2}t$, $J_\parallel=J_s$.

Note that the original electron operator is $c_{i;a\sigma}=\frac{1}{\sqrt{2}}\epsilon_{\sigma \sigma'}\tilde c^\dagger_{i;a \sigma'}$ with $\epsilon_{\sigma \sigma'}$ the $2\times 2$ anti-symmetric tensor. Hence the $\tilde c^\dagger$ operator in the final t-J model should be viewed as a hole creation operator  in the original model. Also the density  operator $\tilde n_{i;a}=1-n_{i;a}$ now has an expectation value $\langle \tilde n_{i;a} \rangle=x$. We note that bilayer $t-J-J_\perp$ model was also proposed to be experimentally relevant in bilayer optical lattice out of equilibrium \cite{bohrdt2022strong,hirthe2023magnetically}.

\textit{Inter-layer paired  superconductor}: 
We focus on the case with large positive $J_\perp$, then the local moments in the two layers (layer 2 and 3 in Fig.~\ref{fig:1}) tend to form on-site spin-singlet. But because
of the $J_K$ term, the entanglement is also shared to the
top and bottom itinerant electrons. 
This mechanism facilitates the formation of interlayer Cooper pairs between electrons in the top and bottom C-layers. 
In the limit $J_\perp,|J_K|\gg t$, the interlayer pairing strength can be estimated by simply calculating the on-site binding energy. 
Ignoring the hopping $t$ and intralayer spin interactions $J_c$,$J_s$, we can analytically solve the on-site local spin Hamiltonian defined at each site. Then, the three states should be considered for every site depending on the number of particles $n_{i;a}$ in the C layers (see Fig.\ref{fig:2}(a)),
\begin{itemize}
    \item $\ket{d}$ : total spin $S=0$ state with $n_{i,t}=n_{i,b}=0$ 
    \item $\ket{f_{a\sigma}}$: total spin $S=\frac{1}{2}$ with $n_{i,a}=1$, $n_{c,\overline{a}}=0$
    \item $\ket{b}$: total spin $S=0$ state with $n_{i,t}=n_{i,b}=1$
\end{itemize}

Here, we used $\overline{t}=b(\overline{b}=t)$. The corresponding energy of the $\ket{d},\ket{b},\ket{f}$ states and their $J_K/J_\perp$ dependence is provided in SM Sec.\textcolor{Red}{III}. Then, the binding energy is defined as 
$E_B=2\epsilon_f-\epsilon_d-\epsilon_b$. In Fig.~\ref{fig:2}(b), we plot the $J_{K}/J_{\perp}$ dependence of the binding energy showing that it is positive for both $J_K<0$ and $J_K>0$, though it is larger in the $J_K>0$ side. Given the positive binding energy, there is an effective inter-layer attractive interaction which leads to inter-layer s-wave superconductor\cite{yang2023strong}.

\begin{figure}[tb]
    \centering
\includegraphics[width=1\linewidth]{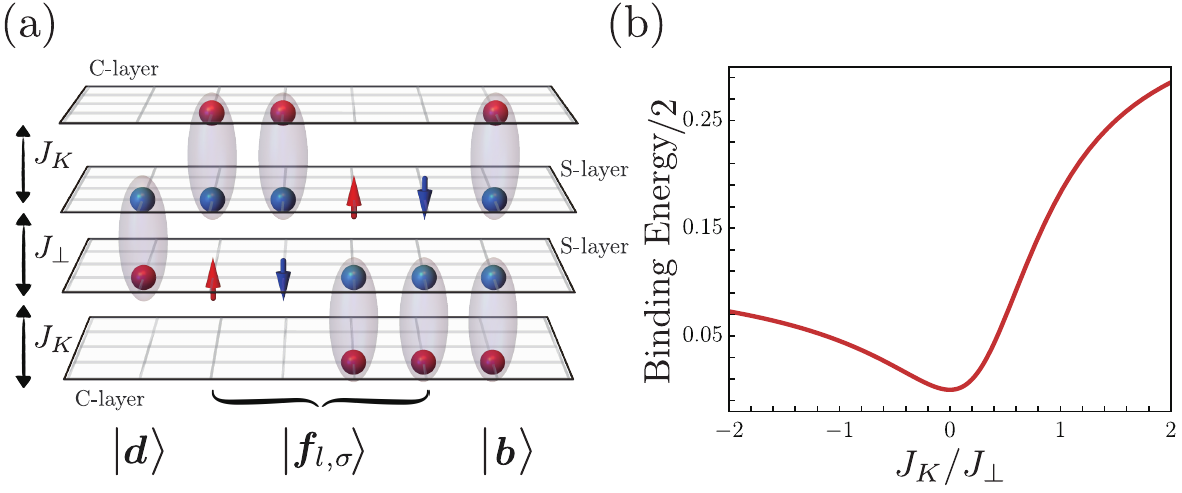}
\caption{
\textbf{Illustration of $\ket{d}$, $\ket{f}$, $\ket{b}$ states and the binding energy energy in $t \rightarrow 0$ limit.} 
(a) The three states are obtained by analytically solving local spin Hamiltonian in term of $J_K$ and $J_\perp$ at each site in $t \rightarrow 0$. For example, this is the configuration in $0<J_{\perp}\ll J_{K}$.
(b) $J_K/J_\perp$ dependence of $E_B/2$ with $E_B=2\epsilon_f-\epsilon_d-\epsilon_b$. The exact function of $E_{B}$ are provided in Eqs. (\ref{eq:onsite}-\ref{eq:binding}) in the supplementary. As shown here, the binding energy is always positive regardless of $J_K/J_{\perp}$. 
}
\label{fig:2}
\end{figure}
 We simulate the double Kondo lattice model in one dimension
using the density matrix renormalization group (DMRG). In Fig.~\ref{fig:3}, we present the finite DMRG results for the model in Eq.(\ref{eq:Ham_optical}) with fixed $L_z=2$, $L_y=1$, where $L_z=2$ correspond to the two coupled Kondo layer.  We set $t=1, J_c=J_s=0.1$. We find a Luther-Emery liquid with power law pairing correlations for the whole range of $|J_K|>0$. In Fig.~\ref{fig:3}(a,b), the $J_K$ dependence of the spin gap is shown for $J_\perp=1$ and $J_\perp=5$. The spin gap, defined as $\Delta_{s}= E(S_z = 1)-E(S_z = 0)$, characterizes the pairing strength. 
 Notably, the pairing in the positive $J_K$ side is a few times larger than the negative $J_K$ side, though the gap at the negative $ J_K$ side is also sizable. The $J_K$ dependence of the spin gap is qualitatively similar to the binding energy  in Fig.~\ref{fig:2} (b). The pair-pair correlation exhibits a power-law behavior with exponent around $1$, as depicted in Fig.~\ref{fig:3}(c). In SM Sec. \textcolor{Red}{IV}, we fit the central charge $c$ to be $c\approx 1$, consistent with a Luther-Emery liquid phase. Although the binding energy is positive, we find that the $\ket{f}$ states (see Fig.~\ref{fig:2}(a))  are still active in  the low energy.   In Fig.~\ref{fig:3} (d), we show that the percentage of the $\ket{f}$ states is sizable even in the large $|J_K|$ limit. Therefore the system is not in the BEC limit although there is a net attractive interaction.
\begin{figure}[tb]
    \centering
\includegraphics[width=1.00\linewidth]{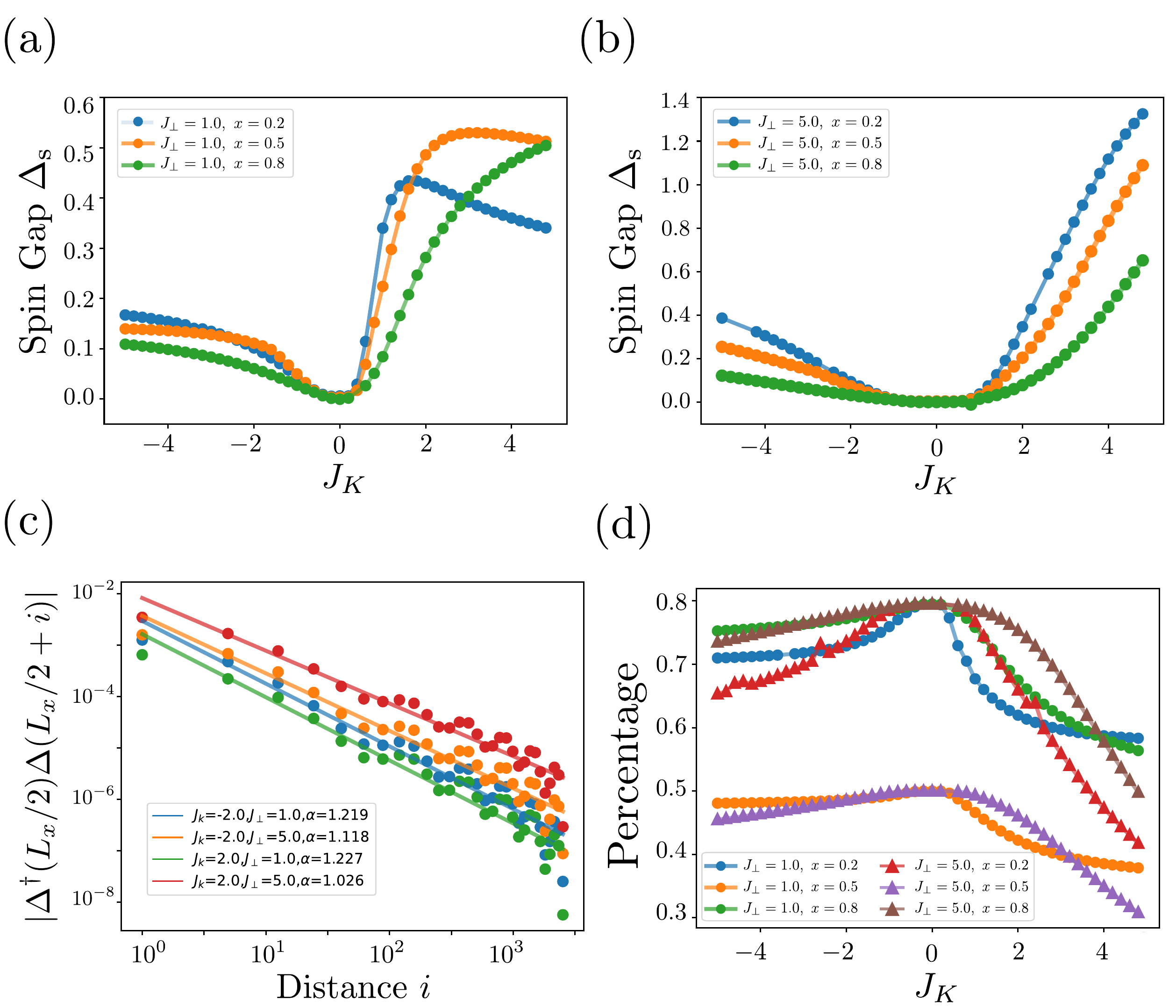}
\caption{\textbf{Finite DMRG results of the double Kondo model, Eq.(\ref{eq:Ham_optical}) in the 1D chain}. 
Here, we use $L_x=60$, $L_y=1$, and a bond dimension $\chi=4000$. We set $t=1, J_c=J_s=0.1$  and only tune $J_K$ and $J_\perp$. 
(a-b) The $J_K$ dependence of the spin gap, $\Delta_S=E(S^z=1)-E(S^z=0)$, at $J_\perp=1$ and $J_\perp=5$, respectively. (c) The pair-pair correlation function for $x=0.2$ in the log-log scale, fitted with the power law relation, $\langle\Delta^\dag(x)\Delta(0)\rangle\sim |x|^{-\alpha}$.
The inter-layer Cooper pairing is defined as $\Delta(i)=\epsilon_{\sigma,\sigma'}\langle c_{i;a,\sigma} c_{i;\overline{a},\sigma'} \rangle$, where $a= t,b$ ($\overline{a}= b,t$) is the layer index, and $\epsilon_{\sigma,\sigma^\prime}$ is an antisymmetric tensor. (d) The $x$ dependence of the   percentage of the $\ket{f}$ state $\frac{n_{f}}{2x}$ ($\frac{n_{f}}{2(2-x)}$) for $x<0.5$ ($x>0.5$).  The percentage of the singlon decreases with $|J_K|$, but is always at order of $10\%$, suggesting that the approach of using a purely bosonic model is not appropriate. 
}
\label{fig:3}
\end{figure}

\textit{Symmetric pseudogap metal}: We have shown that the ground state of the double Kondo lattice model is a superconductor for finite positive $J_\perp$ and finite $|J_K|$. Here we also point out a potential transition in the normal state above $T_c$. Especially, there is a symmetric pseudogap metal without any symmetry breaking in the large $|J_K|$ regime. To capture the potential metallic phase, we perform the standard Abrikosov fermion mean-field theory \cite{Coleman_2015} for the Hamiltonian in Eq.(\ref{eq:Ham_optical}) in the $J_K>0$ side on square lattice. The mean-field Hamiltonian is,
\begin{align}
    H_{MF}=&-t\sum_{a=t,b}\sum_{\langle i,j\rangle} c_{i;a\sigma}^\dag c_{j;a\sigma}+\Phi\sum_{a,i}\left[c_{i;a}^\dag f_{i;a}+h.c.\right]\nonumber\\
+&\Delta\sum_{i,\sigma,\sigma'}\left[\epsilon_{\sigma\sigma^\prime}f_{i;t\sigma}f_{i;b\sigma^\prime}+h.c.\right],
\end{align}
where $c_{i;a\sigma}$ is the free fermion with spin $\sigma$ in layer $a$, while $f_{t/b}$ is introduced for the local moments $\vec{S}_{t/b}=\frac{1}{2}f^\dag_{t/b,\alpha}\vec{\sigma}_{\alpha\beta}f_{t/b,\beta}$, and $\Phi$ is the usual Kondo hybridization between the iterate electron $c$ and the $f$, while $\Delta$ is the pairing of $f_t$ and $f_b$. $\Phi$ is decoupled from the $J_K$ term and $\Delta$ is decoupled from the $J_\perp$ term. The mean field theory is not good at describing the superconductor phase, but can capture the following two different symmetric metallic phases: (I) $\Delta\neq 0$, $\Phi=0$ describes a FL phase with Fermi surface volume $A_{FS}=\frac{1-x}{2}$ in small $J_K$, where the local moments just form rung singlets.   (II) $\Delta=0$, $\Phi\neq0$ corresponds to a symmetric pseudogap metal (sPG) with Fermi surface volume $A_{FS}=-\frac{x}{2}$.  By tuning $J_K$, we can realize the transition from the FL to symmetric pseudogap metal, shown in  Fig.~\ref{fig:4}(a). In Fig.~\ref{fig:4}(b), we show the Fermi surfaces of the FL and sPG respectively. The sPG metal has a hole pocket centered around $\bm{k}=(\pi,\pi)$. Here, the Fermi surface reconstruction is from the hybridization of the itinerant electron and the $f$ band from the local moments.  We do not find superconductor phase because the mean field theory fails to capture the generation of the effective $\tilde J_\perp$ coupling between itinerant electrons. We believe both the FL and sPG phase are not stable to superconductivity at low temperature, but they are expected above $T_c$.

\begin{figure}[t]
    \centering
\includegraphics[width=1\linewidth]{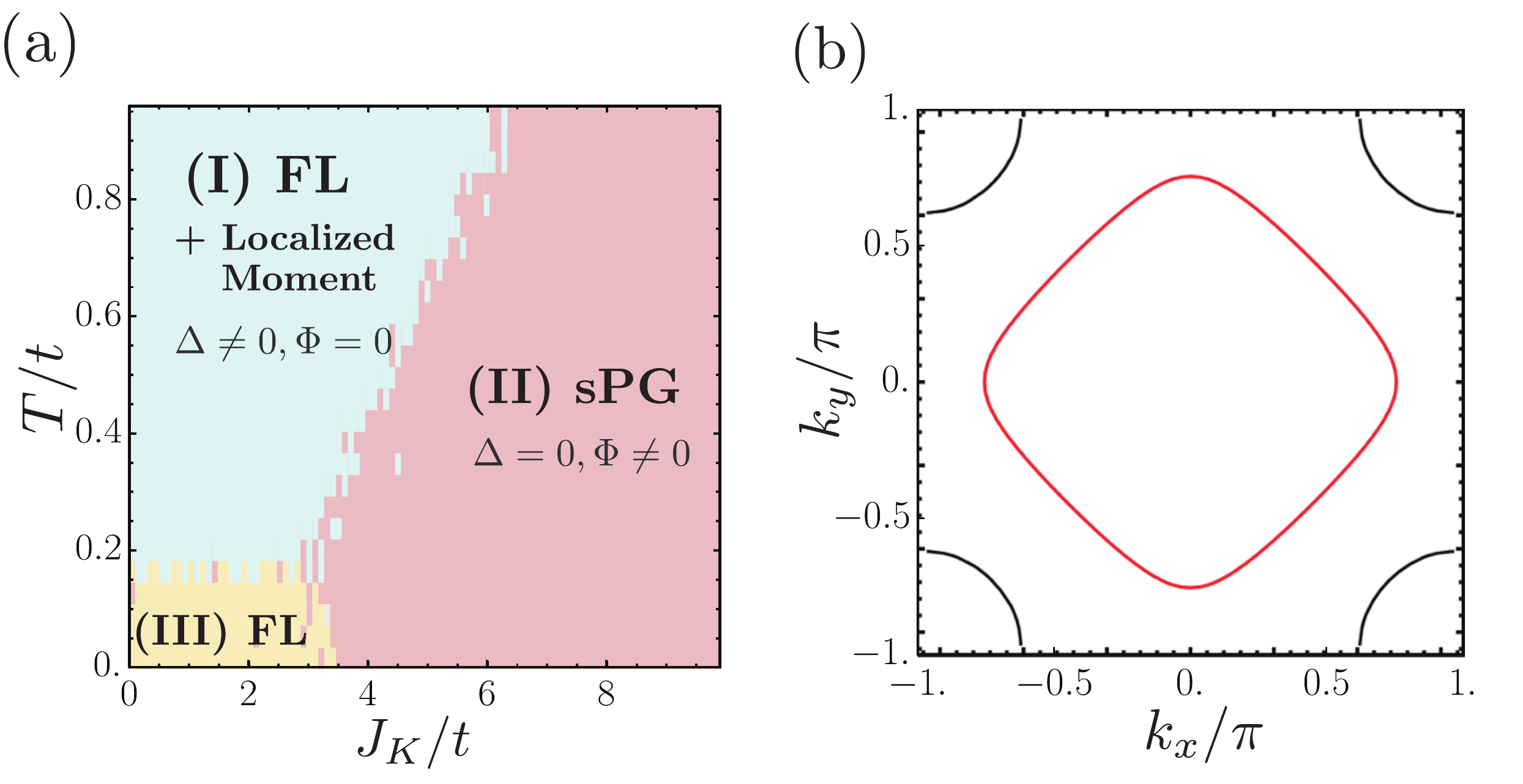}
\caption{\textbf{The Mean-field phase diagram of double kondo model and the illustrated Fermi-surface of the Fermi Liquid(FL) and symmetric pseudogap (sPG).} (a) The Mean-field phase diagram for $J_\perp=1,x=0.2$ is shown as a function of the two dimensionless parameters, $J_K/t$, and $T/t$. 
There appear three different phases: FL with localized moment ($\Delta=\Phi=0$), sPG ($\Delta=0, \Phi\neq 0$), and FL ($\Delta\neq 0, \Phi= 0$), as varied by $J_K$, $T$. 
(b) The Fermi surface of sPG (black) and the FL (red) phases at zero temperature. For illustrations, we choose $J_K/t=4$ for sPG (with mean-field solution $\Phi\approx 1.23,\Delta\approx 0$) and $J_K/t=0$ limit for FL.
}
\label{fig:4}
\end{figure}

\textit{Discussion and Conclusion}: 
In summary, we propose a double Kondo lattice model which unifies the bilayer nickelate and  tetralayer optical lattice system for the $J_K<0$ and $J_K>0$ sides respectively. We demonstrate robust inter-layer superconductivity for both signs of $J_K$. A bilayer one orbital $t-J-J_\perp$ model has been proposed for the bilayer La$_3$Ni$_2$O$_7$\cite{lu2023interlayer}. Our study here indicates the bilayer $t-J$ model is not appropriate for the bilayer nickelate. Instead, one should use the double Kondo lattice model with $J_K<0$ or the type II $t-J$ model\cite{oh2023type} in the $J_K\rightarrow -\infty$ limit. The bilayer $t-J-J_\perp$ model is justified in the large positive $J_K$ regime and may be realized in the tetralayer optical lattice.
Our analysis suggests that the positive $J_K$ side hosts stronger pairing, potentially leading to a high $T_c$ superconductor in realistic cold atom experiments.

In addition to the robust superconductivity, the double Kondo lattice model has the advantage to allow two distinct symmetric Fermi liquids with a Fermi surface volume jump. Especially, a symmetric pseudogap metal with small hole pocket is stabilized at large $J_K$.   This offers an opportunity to study symmetric Kondo breakdown transition\cite{wu2024deconfined} without any symmetry breaking orders. The simple mean field theory predicts a first order transition, but we believe continuous transition is possible once quantum fluctuations are incorporated appropriately, which we hope to explore in the future.  Besides, in the double Kondo lattice model, the symmetric pseudogap metal phase can be conveniently described by the hybridization of the itinerant electron $c$ and the $f$ band from the local moments. As we discussed, the model can be reduced to a one-orbital bilayer $t-J-J_\perp$ model in the large positive $J_K$ limit. We conjecture the same symmetric pseudogap metal phase survives in the bilayer one-orbital model where there is no local moments. In that case the $f$ band should be interpreted as ancilla' degree of freedom proposed by Ref.~\cite{zhang2020pseudogap}. In the future it is interesting apply the ancilla wavefunction\cite{zhang2020pseudogap} to the bilayer $t-J-J_\perp$  model directly to reveal the evolution from a pseudogap metal in the single layer t-J model at small $J_\perp$ limit to the symmetric pseudogap metal at large $J_\perp$ regime.

\textit{Acknowledgement}: 
YHZ thanks Immanuel Bloch, Annabelle Bohrdt and  Fabian Grusdt for insightful discussion. 
H.Oh thanks Kyungtae Kim for the helpful discussion on the cold atom experiments.    
This work was supported by the National Science Foundation under Grant No. DMR-2237031.

\bibliographystyle{apsrev4-1}

\bibliography{ref}

\onecolumngrid
\newpage
\clearpage

\setcounter{equation}{0}
\setcounter{figure}{0}
\setcounter{table}{0}
\setcounter{page}{1}

\maketitle 
\makeatletter
\renewcommand{\theequation}{S\arabic{equation}}
\renewcommand{\thefigure}{S\arabic{figure}}
\renewcommand{\thetable}{S\arabic{table}}

\begin{center}
\vspace{10pt}
\textbf{\large Supplemental Material for \\``Strong pairing and symmetric pseudogap metal in double Kondo lattice model: from nickelate superconductor to tetralayer optical lattice''}
\end{center} 
\begin{center} 
{Hui Yang$^{\ \textcolor{red}{*}}$, Hanbit Oh$^{\ \textcolor{red}{*}}$ and Ya-Hui Zhang$^{\ \textcolor{red}{\dagger}}$}\\
\emph{William H. Miller III Department of Physics and Astronomy, \\
Johns Hopkins University, Baltimore, Maryland, 21218, USA}
\vspace{5pt}
\end{center}

\tableofcontents

\section{I. Realization of Kondo model in the tetra layer optical lattice  }
\label{SS1}
In this section, we derive the double Kondo lattice model starting from the Hubbard model of the tetra-layer optical lattice. 
We start from the optical lattice Fermi-Hubbard model,
\begin{eqnarray}
    H_{\mathrm{opt}}&=& 
    \sum_{l=t,b} H_{l} 
    -\tilde t_{\perp}
    \sum_{\sigma} 
    c^\dagger_{i;2;l}
    c_{i;3;l} +h.c.,\\
H_{t} &= & 
-t\sum_{a=1,2}
\sum_{\langle i,j \rangle;\sigma}
(c^{\dagger}_{i;a;\sigma}
c_{j;a;\sigma}
+h.c.)
-t_{\perp}
\sum_{i;\sigma}
(c^{\dagger}_{i;1;\sigma}
c_{i;2;\sigma}
+h.c.)
\nonumber
\\
&& 
+\Delta \sum_{i}n_{i;1}
+
\frac{U}{2}
\sum_{a;i}
n_{i;a} (n_{i;a}-1)
-\mu 
\sum_{a;i}
n_{a;i},
\nonumber
\\
H_{b} &=& 
\left[ 1\leftrightarrow 4, 2 \leftrightarrow 3
\right]
\nonumber, 
\end{eqnarray}
where $i,a, \sigma$ is the site, layer, spin index, respectively. Here, 
$n_{i;a}=n_{i;a;\uparrow}+n_{i;a;\downarrow}$ is the density at site $i$ and layer $a$.
$\Delta$ is the potential difference between layers 1 and 2 (or 3 and 4), and we here restrict to $0<\Delta<U$. We then fill the layer $2$ and $3$ first so that $n_{i;2}=n_{i;3}=1$. After that, the additional particles prefer to enter layer $1$ and $4$ to reduce the Hubbard $U$. In another word, layer $2$ and $3$ are Mott localized with just spin $1/2$ moment per site. Layer $1$ and $4$ provide itinerant electrons with average density $n_{1}=n_{4}=1-x$.

Assuming $t\ll \Delta <U$, we can  reach the double Kondo model through the standard second-order perturbation theory,
\begin{eqnarray}
    H
    &=&
    -t\sum_{a=t,b}
\sum_{\langle i,j \rangle;\sigma}
(Pc^{\dagger}_{i;a;\sigma}
c_{j;a;\sigma}P
+h.c.)
+J_{c}
\sum_{a=t,b}
\sum_{\langle i,j \rangle }
\vec s_{c,i;a} \cdot \vec s_{c,j;a}
+J_{\perp}\sum_{i}
\vec S_{i;t} \cdot \vec S_{i;b} \notag
\\
&&
+J_{s}
\sum_{a=t,b}
\sum_{\langle i,j \rangle }
\vec S_{i;a} \cdot \vec S_{j;a}
+J_{K} \sum_{a=t,b}\sum_{i}
\vec s_{c,i;a} \cdot \vec S_{i;a},
\label{eq:H_opt}
\end{eqnarray}
with 
\begin{eqnarray*}
    J_c &=& \frac{4t^2}{U}, \quad
    J_{s} =\frac{4t^2}{U}, \quad
    J_{K} = \frac{2t_{\perp}^2}{U-\Delta}+\frac{2t_{\perp}^2}{U+\Delta},
    \quad 
    J_{\perp}=\frac{4\tilde t_\perp^2}{U},
\end{eqnarray*}
where $c_{i;t/b;\sigma}$ (correspond to $c_{i;1/4;\sigma}$ in the optical lattice Hamiltonian) serves as a conduction electron, while $\vec S_{i;t/b}$ (corresponds to $\vec S_{i;2/3}$ in the optical lattice Hamiltonian) is localized moment. 
We have $n_{1}=n_{4}=1-x$ and $n_{2}=n_{3}=1$.

\section{II. Double Kondo lattice model in bilayer nickelate}
The bilayer nickelate La$_3$Ni$_2$O$_7$ is described by a two-orbital Hubbard model on a bilayer square lattice model.
Based on the DFT analysis, it has been shown that two $d$ orbitals, $d_{x^2-y^2}$ and $d_{z^2}$ are relevant to the systems. Since the $d_{z^2}$ orbitals are almost half-filling thus be localized, only providing spin-half moments, while the $d_{x^2-y^2}$ orbitals are mobile electrons. 
The Hamiltonian of the bilayer nickelate is written as, 
\begin{eqnarray}
    H_{\mathrm{nic}}&=& 
    -t\sum_{a=t,b}
\sum_{\langle i,j \rangle;\sigma}
(Pc^{\dagger}_{i;a;\sigma}
c_{j;a;\sigma}P
+h.c.)+
J_c \sum_{a=t,b}\sum_{\langle ij \rangle} \vec s_{i;a;1}\cdot \vec s_{j;a;1}+J_s \sum_{a=t,b}\sum_{\langle ij \rangle} \vec s_{i;a;2}\cdot \vec s_{j;a;2}
+J_\perp \sum_{i} \vec s_{i;t;2}\cdot \vec s_{i;b;2} \notag \\
& &
-2J_H\!\sum_{a=t,b}\sum_{i} 
\left[\vec s_{i;a;1}\cdot \vec s_{i;a;2}+\frac{1}{4}n_{i;a}
\right],\label{eq:H_nic}
\end{eqnarray}
where $c_{i,a,\sigma}$ is an itinerant electron operator of $d_{x^2-y^2}$ orbitals with site $i$, layer $a=t,b$, and spin $\sigma=\uparrow,\downarrow$ indices. Here, the spin operator, $\vec{s}_{i;a;1}=\frac{1}{2}c_{i,a,\sigma}^\dagger\vec{\sigma}_{\sigma,\sigma'}
c_{i,a,\sigma}$, and the density operator $n_{i;a}=\sum_{\sigma} c_{i,a,\sigma}^\dagger c_{i,a,\sigma}$ of $d_{x^2-y^2}$ are introduced. Meanwhile, $\vec{s}_{i;a;2}$ is the localized spin operator of $d_{z^2}$ orbital.  Comparing two Hamiltonians, Eqs.(\ref{eq:H_opt},\ref{eq:H_nic}), we can view the two orbital $d_{x^2-y^2}$ and $d_{z^2}$ defined on the bilayer nickelates as the itinerant electron and the localized moment  in the double Kondo model. The main difference of the two systems is that the effective Kondo coupling derived from the optical lattice is antiferromagnetic with $J_K>0$, while in nickelate it is ferromagnetic Hund's coupling $J_{K}=-2J_H<0$.

As explained in the main-text, if we take $J_{K}\rightarrow -\infty $ limit, Eq.(\ref{eq:H_nic}) can be mapped onto the type-II t-J model. 
This is because large ferromagnetic Hund coupling always enforces the doublon state as the spin-triplet state, described by the spin-1 operators. Hence, the the Hilbert space at each site is restricted as a combination of these spin-1 moments of the doublon state and the spin-1/2 moment from the singlon state.  
The spin operator of the singlon state are $ \vec S_{i;a}=\frac{1}{2}\sum_{\sigma \sigma'} \ket{\sigma}_{i,a} \vec \sigma_{\sigma \sigma'}\bra{\sigma'}_{i,a}$ with $a=t,b$ and $\vec \sigma$ as the Pauli matrices. The spin operators for doublon states are written as $ \vec T_{i,a}=\sum_{\alpha,\beta=-1,0,1} \vec T_{\alpha \beta}\ket{t}_{\alpha,i,a} \bra{t}_{\beta,i,a}$. Here,  we have 
 \begin{eqnarray*}
     T_x=\frac{1}{\sqrt{2}}\begin{pmatrix} 0 & 1 & 0 \\ 1 & 0 & 1 \\ 0 & 1 & 0 \end{pmatrix} , 
     \quad 
     T_y=\frac{1}{\sqrt{2}}\begin{pmatrix} 0 & -i & 0 \\ i & 0 & -i \\ 0 & i & 0 \end{pmatrix}, 
     \quad 
      T_z=\begin{pmatrix} 1 & 0 & 0 \\ 0& 0 & 0 \\ 0 & 0 & -1 \end{pmatrix}, 
     \quad 
 \end{eqnarray*}
in the $\ket{t}_1,\ket{t}_0,\ket{t}_{-1}$ basis. 
The doublon states defined at the $a=t,b$ layer are given as $\ket{t}_{1,i,a}=d^\dagger_{i,a,1,\uparrow}d^\dagger_{i,a,2,\uparrow}\ket{G}$, $\ket{t}_{0,i,a}=\frac{1}{\sqrt{2}}(d^\dagger_{i,a,1,\uparrow}d^\dagger_{i,a,2,\downarrow}
+d^\dagger_{i,a,1,\downarrow}d^\dagger_{i,a,2,\uparrow})\ket{G}$ and $\ket{t}_{-1,i,a}=d^\dagger_{i,a,1,\downarrow}d^\dagger_{i,a,2,\downarrow}\ket{G}$.
Note that $d^\dagger_{i,a,1,\sigma}=c^\dagger_{i,a,\sigma}$ is an electron operator of the $d_{x^2-y^2}$ orbital, and $d^\dagger_{i,a,2,\sigma}$ is that of $d_{z^2}$ orbital. $\ket{G}$ is the vacuum.
Finally, the resultant Hamiltonian is given as the type-II t-J Hamiltonian,
\begin{eqnarray}
     H&=&   
     -t\sum_{a=t,b}
\sum_{\langle i,j \rangle;\sigma}
(Pc^{\dagger}_{i;a;\sigma}
c_{j;a;\sigma}P
+h.c.),
     \\
   &&+  \sum_{a=t,b} \sum_{\langle ij \rangle}
   \left[
   J^{ss}_\parallel
\vec S_{i;a}\cdot \vec S_{j;a}
+J^{sd}_\parallel 
      (\vec S_{i;a}\cdot \vec T_{j;a} +\vec T_{i;a}\cdot \vec S_{j;a})
      +
      J^{dd}_\parallel
\vec T_{i;a}\cdot \vec T_{j;a} \notag 
      \right]
      \\
&&+ \sum_{i} \left[J^{ss}_\perp\vec S_{i;t}\cdot \vec S_{i;b}
+
J^{sd}_\perp 
(\vec S_{i;t}\cdot \vec T_{i;b}+\vec T_{i;t}\cdot \vec S_{i;b})
+J ^{dd}_\perp  \vec T_{i;t}\cdot \vec T_{i;b} \right],\notag
\label{eq:type_II_t_J}
\end{eqnarray}
with $J^{ss}_{\perp}=2J^{sd}_{\perp}=4J^{dd}_{\perp}=J_{\perp}$, and 
$J^{ss}_{\parallel}=J_s$, $J^{sd}_{\parallel}=J_s/2$, $J^{dd}_{\parallel}=J_c/4+J_s/4$.

\section{III. Details on $\ket{d},\ket{f},\ket{b}$ states and the binding energy in $t\rightarrow0 $ limit }
Starting from the double Kondo model, we take the limit of $t\rightarrow 0$. In this limit, we can ignore the hopping and intra-layer spin exchange interaction, thus it is enough to solve the local spin interaction defined at every site $i$,
\begin{align}
 H=&
J_{K}
\sum_{a=t,b} \vec s_{c;a} \cdot \vec S_{a}+J_{\perp}
\vec S_{t}\cdot \vec S_{b},
\label{eq:Ham_local}
\end{align}
where we omit the site index $i$ here. 
Depending on the particle number of C-layer,$n_{C}=\sum_{a,\sigma}c^{\dagger}_{a;\sigma}c_{a;\sigma}$ at each site, the three states can describe the low energy physics. 
The wave function is given as 
\begin{eqnarray*}
  n_C=0: \quad  |d\rangle &=&  \frac{1}{\sqrt{2}}\left[
   |\uparrow\downarrow\rangle-|\downarrow\uparrow\rangle
   \right]_{2,3},
   \end{eqnarray*}
\begin{eqnarray*}
n_C=1: \quad       |f_{t,\sigma}\rangle&
  \sim&
  \left[
|\sigma\sigma\overline{\sigma}\rangle
-
\left(\tilde{J}_{K}+\alpha(\tilde{J}_{K})\right)
|\sigma\overline{\sigma}\sigma\rangle
+
\left(\tilde{J}_{K}+\alpha(\tilde{J}_{K})-1\right)
|\overline{\sigma}\sigma\sigma\rangle
\right]_{1,2,3}
   \\
|f_{b,\sigma}\rangle&
  \sim&
  \left[
|\sigma\sigma\overline{\sigma}\rangle
-
\left(\tilde{J}_{K}+\alpha(\tilde{J}_{K})\right)
|\sigma\overline{\sigma}\sigma\rangle
+
\left(\tilde{J}_{K}+\alpha(\tilde{J}_{K})-1\right)
|\overline{\sigma}\sigma\sigma\rangle
\right]_{4,3,2}
\end{eqnarray*}
\begin{eqnarray*}
  n_C=2: \quad |b\rangle &
  \sim&
  \left[
 |\uparrow \uparrow \downarrow\downarrow \rangle
 + |\downarrow \downarrow\uparrow \uparrow \rangle
-(2\tilde{J}_{K}+\beta(\tilde{J}_{K}))  
( |\uparrow \downarrow \uparrow\downarrow \rangle
 + |\downarrow \uparrow\downarrow \uparrow \rangle)
+(2r+\beta(\tilde{J}_{K})-1)  
( |\uparrow \downarrow \downarrow\uparrow \rangle
 + |\downarrow \uparrow\uparrow \downarrow \rangle)
\right]_{1,2,3,4}.
\end{eqnarray*}
The subscript $1,2,3,4$ denotes the layer index defined in Fig.\ref{fig:1}. We have defined $ \alpha(r) =\sqrt{1 -r+ r^2} , 
   \beta(r) =\sqrt{1 -2r+ 4r^2}$
with $\tilde{J}_{K}=J_{K}/J_{\perp}$. 
The corresponding energies of $\ket{d},\ket{f},\ket{b}$ states are given by 
\begin{eqnarray}
\epsilon_d&=&-\frac{3}{4}J_\perp,\nonumber \\
   \epsilon_f&=&\frac{J_\perp}{4}\left[-\tilde{J}_K
   -1-2\alpha(\tilde{J}_K)\right],    \label{eq:onsite}\\
   \epsilon_b&=&\frac{J_\perp}{4}\left[-2\tilde{J}_K-1-2\beta(\tilde{J}_K)\right]\nonumber
\end{eqnarray}
thus the binding energy $E_{B}=2\epsilon_{f}-\epsilon_d-\epsilon_b$ becomes, 
\begin{eqnarray}
E_{B}&=&
   \frac{J_{\perp}}{2} [1 + 
   \beta(\tilde{J}_K)
   -2 \alpha(\tilde{J}_K)].
   \label{eq:binding}
\end{eqnarray}

\section{IV. DMRG details and detailed DMRG results for the double Kondo model}
\begin{figure}[h]
    \centering
\includegraphics[width=0.7\linewidth]{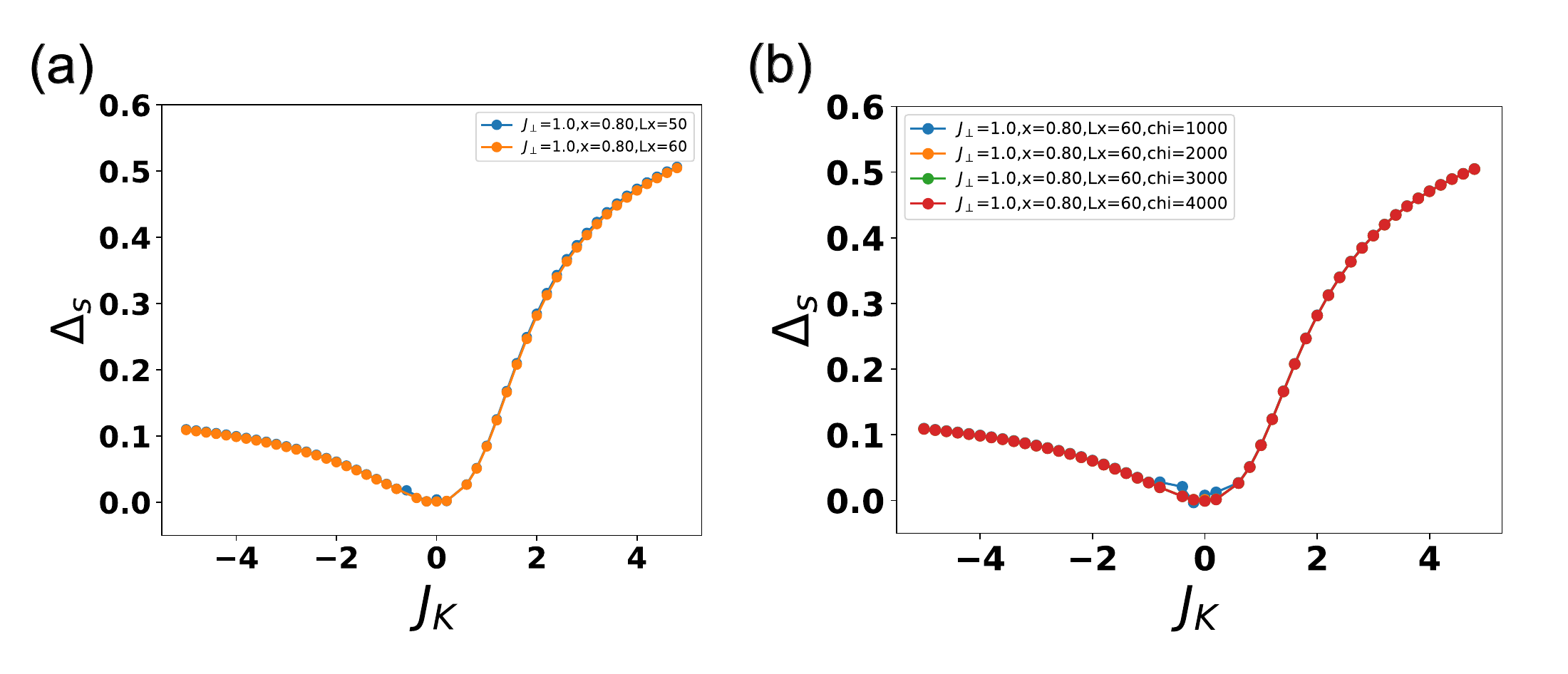}
\caption{(a) the spin gap $\Delta_S$ for different $L_x$ with $J_s=J_c=0.1$, $J_\perp=1$, $x=0.8$. (b) the spin gap $\Delta_S$ for different bond dimension $\chi=1000,2000,3000,4000$ with $J_s=J_c=0.1$, $J_\perp=1$, $x=0.8$.
}
\label{fig:convergence}
\end{figure}

In our DMRG simulation,  we split the double Kondo model site to two sites, corresponding to the top and bottom layer separately. Each site has $3\times 2=6$ states.  In this notation the 1D double Kondo lattice model is the same as the conventional Kondo lattice model on a two-leg ladder, but with an inter-layer $J_\perp$ coupling. The total particle number (in fact the particle in each layer is conserved) and the total $S_z$ are conserved in this model. Here we provide more detailed DMRG results of the double Kondo model. The spin gaps for different $L_x$ are shown in Fig.~\ref{fig:convergence}(a), and we can see the result is convergent for different system size. In Fig.~\ref{fig:convergence}(b), we can see the spin gap is convergent with bond dimension. Bond dimension $\chi=4000$ should be enough.

\begin{figure}[h]
    \centering
\includegraphics[width=0.9\linewidth]{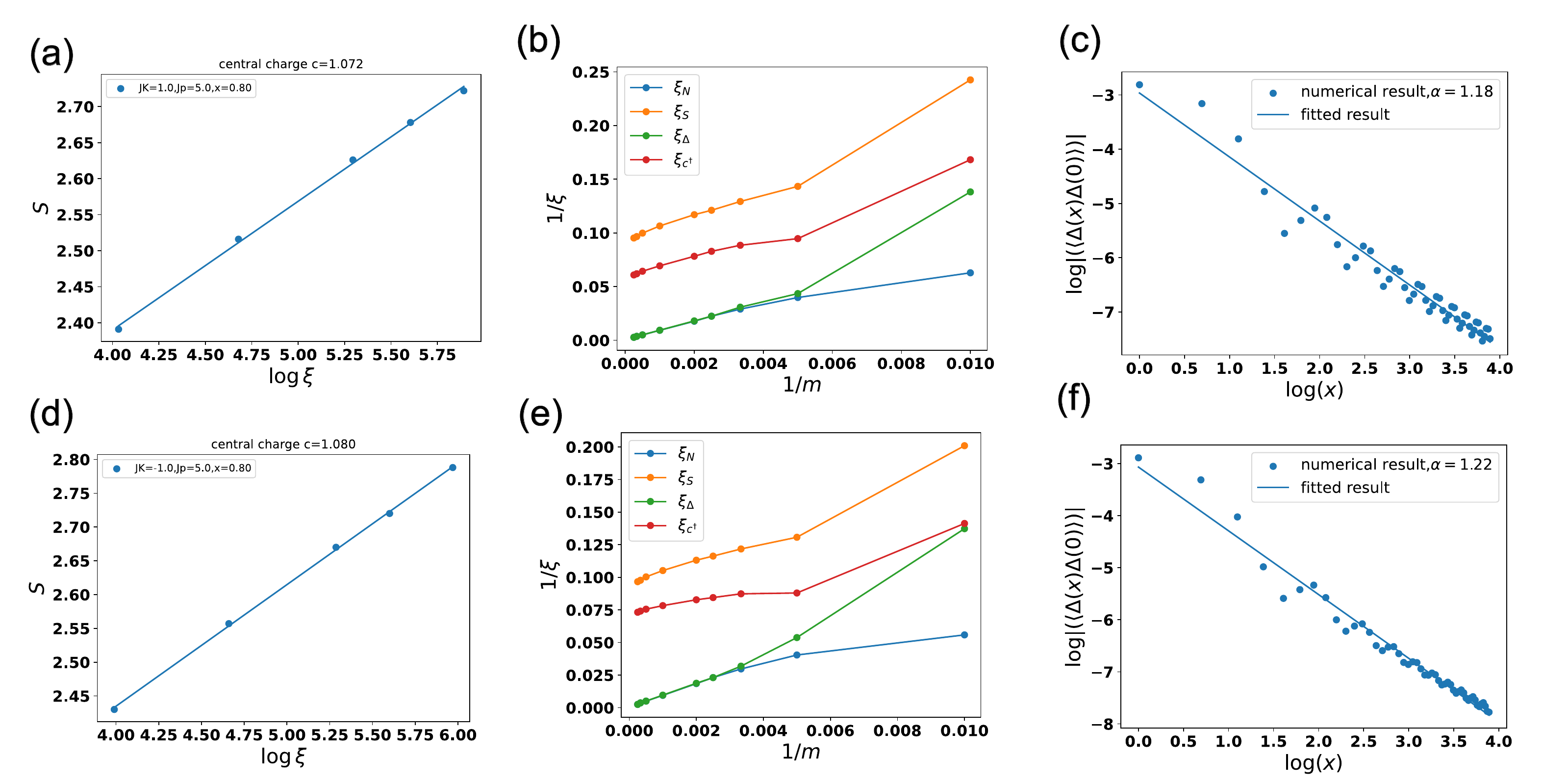}
\caption{(a) and (d) the central charge of $J_K=1$ and $J_K=-1$, respectively. We fit the central charge by the relation $S=\frac{c}{6}\log\xi$, where $S$ is the entanglement entropy, $\xi$ is the correlation length and $c$ is the central charge. (b) and (e) the inverse correlation length of $J_K=1$ and $J_K=-1$, respectively. In our DMRG calculation, there are two quantum numbers, total $N$ and total $S_z$. We can get the correlation length of different operators from the transfer matrix technique in  different symmetry sectors $(\delta N,\delta S)$, with $N$, $S$, $\Delta$, $c^\dag$ corresponding to $(0,0)$, $(0,1)$, $(2,0)$, $(1,\frac{1}{2})$. (c) and (f) the pair-pair correlation function of $J_K=1$ and $J_K=-1$, respectively. We fit the pair-pair correlation function by the power-law relation $|\langle \Delta^\dag(x)\Delta(0)\rangle|=\frac{A}{x^\alpha}$, with $\Delta(x)=\epsilon_{\sigma\sigma^\prime}c^\dag_{t\sigma}(x)c^\dag_{b\sigma^\prime}(x)$.
}
\label{fig:idmrg}
\end{figure}

In Fig.~\ref{fig:idmrg}, we show the central charge (a) and (d), the correlation length (b) and (e), the pair-pair correlation function (c) and (f), for both positive and negative $J_K$. In Fig.~\ref{fig:idmrg} (a) and (d), we fit the central charge from the entanglement entropy $S$ and the correlation length $\xi$ by the relation $S=\frac{c}{6}\log\xi$, where $c$ is the central charge, and we can find the central charge is $c=1$ within error. From the correlation length in Fig.~\ref{fig:idmrg} (b) and (e), we can see the spin degree of freedom is gapped, while the charge degree freedom remains gapless, which is consistent with the Luther-Emery liquid. In Fig.~\ref{fig:idmrg}(c) and (f), we fit the pair-pair correlation function by the power-law relation $|\langle \Delta^\dag(x)\Delta(0)\rangle|=\frac{A}{x^\alpha}$, where $\Delta(x)=\epsilon_{\sigma\sigma^\prime}c^\dag_{t\sigma}(x)c^\dag_{b\sigma^\prime}(x)$, and we find the pair-pair correlation function shows a power-law behavior, which is another property in the Luther-Emery liquid.

\section{V. Mean-field solution of the double-Kondo model}
\begin{figure}[h]
    \centering
\includegraphics[width=1.00\linewidth]{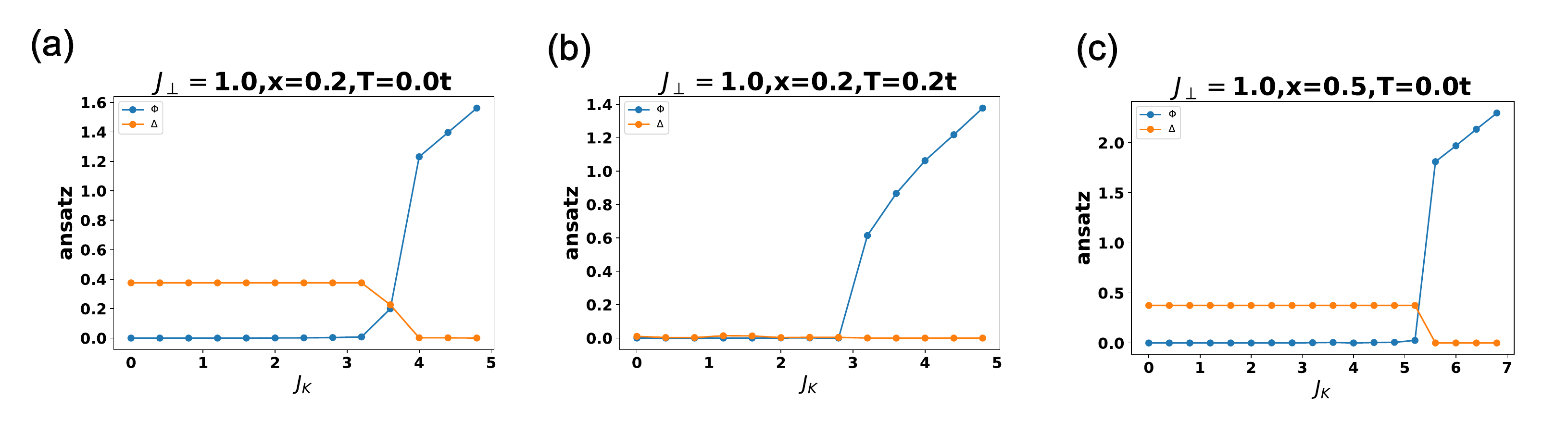}
\caption{The mean-field solutions for $t=1,J_\perp=1$, (a) $x=0.2, T=0$, (b) $x=0.2, T=0.2t$, (c) $x=0.5, T=0$.
}
\label{fig:MF}
\end{figure}

Here we consider the double Kondo model with no intralayer spin interaction,
\begin{align}
    H=-t\sum_{a=t,b}\sum_{\langle ij\rangle}c_{i;a\sigma}^\dag c_{i;a\sigma}+J_\perp\sum_{i}\vec{S}_{i;t}\cdot\vec{S}_{i;b}+J_K\sum_i(\vec{s}_{e;i;t}\cdot\vec{S}_{i;t}+\vec{s}_{e,i;b}\cdot\vec{S}_{i;b}),
\end{align}
We can write spin moments in terms of the spinons, $\vec{S}_{i;a}=\frac{1}{2}f_{i;a}^\dag\vec{\sigma}f_{i;a}$. Introducing the following two mean-field parameters, $\Phi$ and $\Delta$, where $\Phi$ is the hybridization between the itinerant electron and the spinon, while $\Delta$ is the interlayer pairing of of the spinons in the top and bottom layer. In terms of the mean-field parameters, we arrive at the following mean-field Hamiltonian,
\begin{align}
    H_{MF}=-t\sum_{a=1,2}\sum_{\langle ij\rangle} c_{i;a\sigma}^\dag c_{i;a\sigma}+\Phi\sum_{i,l}c_{i;l;\sigma}^\dag f_{i;l;\sigma}+h.c.+\Delta\sum_{i}(\epsilon_{\sigma\sigma^\prime}f_{i;t\sigma}f_{i;b\sigma^\prime}+h.c.),
\end{align}
where the mean-field ansatz are
\begin{align}
    \Phi&=-\frac{3}{8}J_K\frac{1}{2}\sum_{l,\sigma}\langle c_{i;l\sigma}^\dag f_{i;l\sigma}\rangle,\\
    \Delta&=\frac{3}{8}J_\perp\epsilon_{\sigma\sigma^\prime}\langle f_{i;t\sigma}^\dag f_{i;b\sigma^\prime}^\dag\rangle,
\end{align}
as well as two constraint
\begin{align}
    1-x&=\frac{1}{N}\sum_{i}\frac{1}{2}\sum_{l,\sigma}\langle c^\dag_{i;l\sigma}c_{i;l\sigma}\rangle,\\
    1&=\frac{1}{N}\sum_{i}\frac{1}{2}\sum_{l,\sigma}\langle f^\dag_{i;l\sigma}f_{i;l\sigma}\rangle
\end{align}
Solving the mean-field equation self-consistently, we can get the following phases. (I) the spinons are paired and there is no hybridization, corresponding to the Fermi liquid phase with $\Delta\neq 0$ and $\Phi=0$. (II) the spinons are not paired but the hybridization is non-zero, corresponding to the symmetric pseudogap phase with $\Delta=0$ and $\Phi\neq 0$. (III) the spinons are not paired and the hybridization is $0$, corresponding to the phase with the coexistence of itinerant electrons and fluctuating local moments at high temperature. The full phase diagram is shown in in Fig.~\textcolor{Red}{4}(a) in the main text. In Fig.~\ref{fig:MF}, we show some line cuts of the phase diagram.

\begin{figure}[h]
    \centering
\includegraphics[width=.65\linewidth]{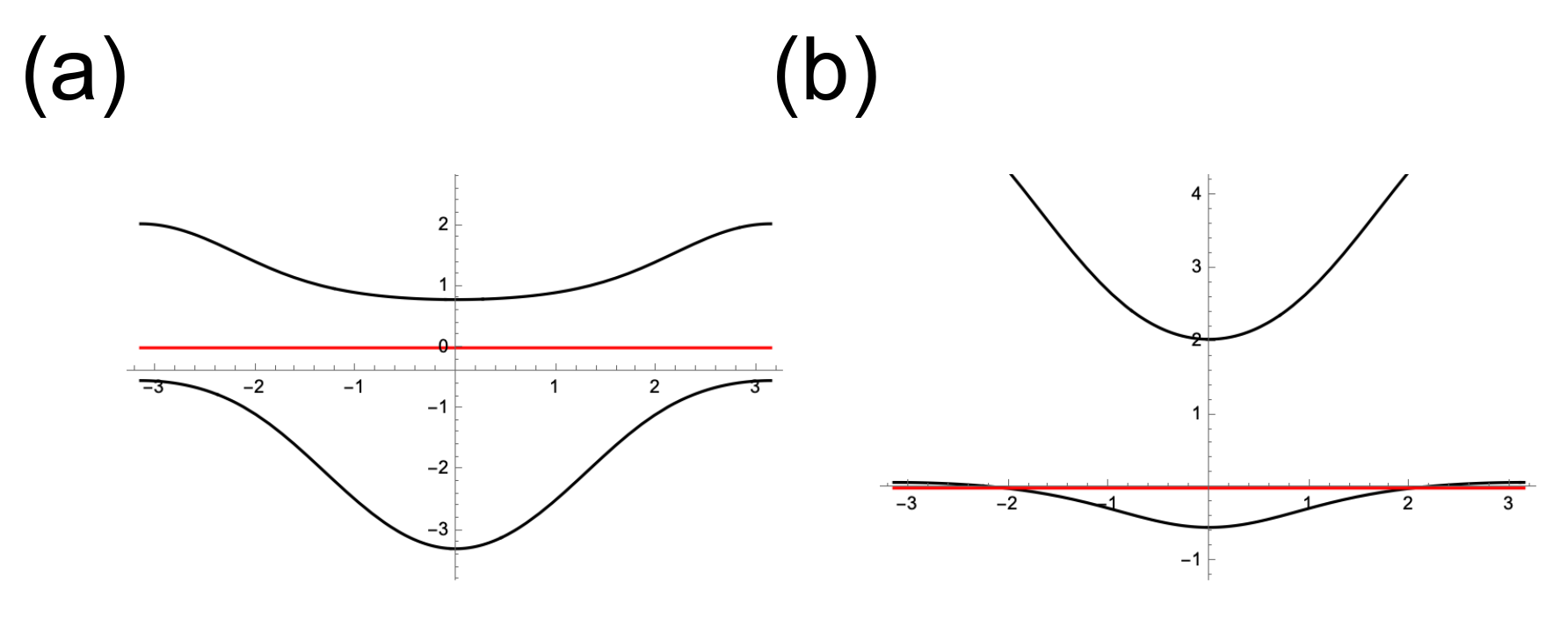}
\caption{The dispersion from the mean-field solution with $J_\perp=1, J_K=4,x=0.2,\Phi\approx 1.23,\Delta\approx 0$. (a) the dispersion along $k_y=0$ and (b) the dispersion along $k_y=\pi$ of the sPG. In the plot the red line corresponds to the Fermi energy.
}
\label{fig:dispersion}
\end{figure}

The dispersion of the symmetric pseudogap metal is shown in Fig.~\ref{fig:dispersion} from the mean-field solution with $J_\perp=1, J_K=4,x=0.2,\Phi\approx 1.23,\Delta\approx 0$. Fig.~\ref{fig:dispersion}(a) and (b) correspond to the dispersion along $k_y=0$ and $k_y=\pi$, respectively. The red line in Fig.~\ref{fig:dispersion} is the Fermi energy. From Fig.~\ref{fig:dispersion}(b), we can see the Fermi pocket centered around ${\bf k}=(\pi,\pi)$.

\end{document}